\documentclass[a4paper, 11pt]{article}
\usepackage{epsfig, amssymb, amsbsy}

\newtheorem{Thm}{Theorem}
\newtheorem{Def}{Definition}

\title{A Note on Mathematical Modelling of Practical Multicampaign Assignment and Its Computational Complexity}

\author{{\small Yong-Hyuk Kim$^\dag$ ~~and~~ Yourim Yoon$^\ddag$}\\ \\
{\small $\dag$ Department of Computer Science \& Engineering, Kwangwoon University}\\
{\small 447-1, Wolgye-dong, Nowon-gu, Seoul, 139-701, Korea}\\
{\small Homepage: {\tt http://soar.snu.ac.kr/\~{ }yhdfly}}\\
{\small Email: {\tt yhdfly@kw.ac.kr}}\\
{\small $\ddag$ School of Computer Science \& Engineering, Seoul National
University}\\
{\small Sillim-dong, Gwanak-gu, Seoul, 151-744, Korea}\\
{\small Email: {\tt yryoon@soar.snu.ac.kr}}
}

\begin{document}
\baselineskip=2.0eM
\maketitle
\begin{abstract}
Within personalized marketing, a recommendation issue known as {\it multicampaign assignment} is 
to overcome a critical problem, known as the multiple recommendation problem which occurs
when running several personalized campaigns simultaneously.
This paper mainly deals with the hardness of multicampaign assignment,
which is treated as a very challenging problem in marketing.
The objective in this problem is to find a customer-campaign matrix which
maximizes the effectiveness of multiple campaigns under some constraints.
We present a realistic response suppression function, which is
designed to be more practical, and explain how this can be learned from historical data.
Moreover, we provide a proof that this more realistic version of the problem is NP-hard,
thus justifying to use of heuristics presented in previous work.\\
{\bf Keywords}: Computational complexity, personalized marketing, multicampaign assignment.
\end{abstract}

\section{Introduction}
CRM (customer relationship management) \cite{Dy01} is important 
in obtaining and maintaining loyal customers.
To maximize income and customer satisfaction, 
firms try to provide personalized services for customers. 
As personalized campaigns are frequently performed,
several campaigns often happen to run at the same time or in a short period of time.
It is often the case that an attractive customer
for a specific campaign tends to be attractive for other campaigns.
If the other campaigns are not considered when assigning customers to a specific campaign,
some customers may be bombarded by a considerable number of campaigns \cite{KM05, KM06}.
This problem is called the {\em multiple recommendation problem}.
As the number of recommendations to a particular customer increases,
that customer's interest in all campaigns decreases \cite{BL00}.
Eventually the rate of customer response for campaigns will drop.
This clearly lowers the targeting efficiency as well as customer satisfaction,
which leads to diminished customer loyalty.
Unfortunately, most traditional approaches have focused 
on the effectiveness of individual campaigns and
did not consider the problem of multiple recommendations.
In the situation where several campaigns are conducted within a short period of time,
we need to find the optimal assignment of campaigns to customers
considering the recommendations in other campaigns.

The {\em multicampaign assignment problem} (MCAP) is a complex assignment problem
in which each of $n$ customers is assigned or not to each of $k$ campaigns.
The goal is to find a set of assignments such that
the effect of the campaigns is maximized under some constraints.
The main difference between this problem and with independent campaigns is that
the customer response for campaigns is influenced by multiple recommendations.
In previous work \cite{KM04, KM05, KM06, KM06-2},
we defined MCAP and proposed three types of methods for it.
We showed that one can solve MCAP to optimality by dynamic programming.
Although the dynamic programming algorithm guarantees optimal solutions,
it becomes intractable for large scale problems.
We thus proposed heuristic methods that not only have practical time complexity
but also showed good performance.
There was also a recent study using particle swarm optimization to solve the
multicampaign assignment problem \cite{DehC08}.
However, since they are heuristics, they do not guarantee optimal solutions.
Furthermore, although many approaches for MCAP have been proposed so far,
nobody has yet shown whether or not the problem is
computationally intractable.

In this paper, we suggest an advanced version
of the multicampaign assignment problem that reflects a more realistic situation
and to which all previous heuristics can be applied as they are, and then prove
that it is NP-complete, using a novel technique.
The remainder of this paper is organized as follows.
In Section~\ref{sec:mcap}, we briefly review the multicampaign assignment problem and its optimal solver.
We design a reasonable response suppression function 
and then present a practical version of the multicampaign assignment problem in Section~\ref{sec:gen}.
We provide the NP-completeness of multicampaign assignment in Section~\ref{sec:npc}.
Finally, we give our concluding remarks in Section~\ref{sec:con}.

\section{Preliminaries\label{sec:mcap}}
\subsection{Multicampaign Assignment}
The basic idea of the multicampaign assignment problem is from \cite{KM06}.
Let $n$ be the number of customers and $k$ be the number of campaigns.
In the following, we describe input, output, constraints, and objective for MCAP.

\noindent 
{\bf Input}: Each campaign has a weight. Each customer has a preference for each
campaign and there is a response suppression
function that gives the rate at which a customers preference for all
campaigns decreases as the number of campaigns
assigned to that customer increases.\\
-- $\boldsymbol{w} = (w_1, w_2, \ldots, w_k) \in \mathbb{Z}_+^k$: the campaign weight vector 
 ($w_j > 0$ is the weight of campaign $j$).\\
-- $P = (p_{ij})$: the preference matrix. Each element $p_{ij} \in \mathbb{Z}_+ \cup \{0\}$ 
is the preference value of customer $i$ for campaign $j$.\\
-- $r : \mathbb{Z}_+ \cup \{0\} \longrightarrow [0, 1]$:
the response suppression function with respect to the number of recommendations
(for convenience, we assume $r(0) = 0$).
The personalized extension of the response suppression function
was considered in \cite{DehC08,KM06-2}.
We will deal with this issue again in Section~\ref{sec:R}.

\noindent
If $h_i$ is the number of recommendations for customer $i$,
the actual preference of customer $i$ for campaign $j$ becomes $r(h_i) p_{ij}$.

\noindent 
{\bf Constraints}:
Let $b^j$ be the upper bound of recommendations for campaign $j$,
and $b_j$ be the lower bound of recommendations for campaign $j$.
Then the actual number of recommendations in campaign $j$ is constrained to lie between $b_j$ and $b^j$.\\
-- $\boldsymbol{b^*} = (b^1, b^2, \ldots, b^k) \in \mathbb{Z}_+^k$: the upper bound capacity constraint vector \\
-- $\boldsymbol{b_*} = (b_1, b_2, \ldots, b_k) \in \mathbb{Z}_+^k$: the lower bound capacity constraint vector \\
These values are predefined largely depending on the budget of each campaign.
The required cost of campaign $j$ will be between $b_j \cdot c_j$ and $b^j \cdot c_j$,
where $c_j$ is the unit recommendation cost for campaign $j$.

\noindent 
{\bf Output}:
The output is an $n \times k$ binary campaign assignment matrix $M = (m_{ij})$ in which
$m_{ij}$ indicates whether or not campaign $j$ is assigned to customer $i$.

\noindent 
{\bf Objective}:
The {\em campaign preference} for campaign $j$ is defined to be
the actual preference sum of recommended customers for campaign $j$
as follows: $\sum_{i=1}^n r(h_i) p_{ij} m_{ij}$,
where $h_i = \sum_{j=1}^k m_{ij}$.
The fitness $F(M)$ of a campaign assignment matrix $M = (m_{ij})$ is
the weighted sum of campaign preferences.
$$F(M) = \sum_{j=1}^k \left(w_j \sum_{i=1}^n r(h_i) p_{ij} m_{ij}\right)_.$$
Note that computing $F(M)$ takes $O(nk)$ time.

\noindent
The objective is to find a binary matrix $M = (m_{ij})$ that maximizes the fitness $F(M)$.

\subsection{Getting Preferences}
The preferences for a campaign can be obtained using existing methods
such as collaborative filtering \cite{RISBR94}.
Collaborative filtering (CF) is widely used to predict customer preferences 
for campaigns \cite{BL97}. In particular, CF is fast and simple \cite{GRGP01},
so it is popular for personalization in e-commerce \cite{Jonathan99, Joseph97}.
There have been a number of customer-preference estimation methods
based on CF \cite{AWWY99, Gre97, HKTR04, HZC07, LLX05, SKKR01, SM95}.
Nearest neighbor CF algorithms, e.g, \cite{SM95},
are based on computing the distance
between customers based on their preference history. Predictions of how much a
customer will like a campaign are computed by taking the weighted average of
the opinions of a set of nearest neighbors for that campaign. Neighbors who
have expressed no opinion on the campaign are ignored.
This type of CF algorithm has been successfully used in previous multicampaign studies
\cite{KM06, KM06-2}.

\subsection{Optimal Algorithm}
We can find the optimal campaign assignment matrix of MCAP
using dynamic programming \cite{KM06}. The algorithm requires
$O(nk \Pi_{j = 1}^k b^j)$ space and takes $O(nk2^k \Pi_{j = 1}^k b^j)$ time.
The dynamic programming approach tries to solve all subproblems of the form
``what is the best assignment for the capacity $(b_1', b_2', \ldots, b_k')$, using just the
first $m$ customers,'' for all $0 \le b_1' \le n, \ldots, 0 \le b_k' \le n$, and
$0 \le m \le n$.
If $k$ is a fixed number, this is a polynomial-time algorithm.
However, when $k$ is not small and almost all $b^j$s are $\Omega(n)$, it is nearly intractable.
A practical version of MCAP, which is described in the next section, can also be solved by
this dynamic programming algorithm with the same complexity.

\section{Practical Version of Multicampaign Assignment\label{sec:gen}}

\subsection{Realistic Response Suppression Function\label{sec:R}}
In the case of multiple campaign recommendations,
the customer response rate changes with recommendation patterns.
We introduced the response suppression functions 
for the response rate with multiple recommendations \cite{KM06}.
But, these functions are very simple since
not only are they artificially generated but
the same function is also applied to all customers.
Recently, this problem was weightily mentioned in \cite{KM06-2} and
furthermore a generalized Gaussian response suppression function that differs among
customer classes was introduced in \cite{DehC08}.
We design a response suppression function that is
more suitable for the practical campaign situation, as mentioned in prior work \cite{DehC08,KM06-2}.

We also assume that each customer can have its own response suppression function
since the degree of response suppression may vary from customer to
customer \cite{BL00, LLX05}.\\
-- $r_i : \mathbb{Z}_+ \cup \{0\} \longrightarrow [0, 1]$:
the response suppression function of customer $i$ with respect to the number of recommendations
(for convenience, we assume $r_i(0) = 0$ and 
$\boldsymbol{r} = (r_1, r_2, \ldots, r_n)$).
We now describe how such a response suppression function can be learned from historical data

Assuming that the response suppression rate of a customer varies as his/her inclination,
we classify the given (training) customers into a number of categories
according to their profiles and preference history.
This can be done by using existing data mining techniques \cite{BL97,BL00,JMF99}.
We are to give the same response suppression function to all the customers 
belonging to the same category.
Each customer has their preference for each campaign estimated by, e.g., CF.
We assume that the prediction error is not large.

We perform each campaign independently to obtain multicampaign data.
If two customers belong to the same category but showed different responses,
we can consider that it is because of response suppression.
We estimate the response suppression by using function approximation to find a
function that satisfies as many of the following conditions as possible.\\
{\bf Condition}:
For each customer pair $(i, j)$ in $C$ and each campaign $k$,
if customer $i$ responded to campaign $k$ and customer $j$ did not,
$$p_{ik} \cdot r_C(h_i) >  p_{jk} \cdot r_C (h_j),$$
where $p_{lk}$ is the preference of customer $l$ for campaign $k$, $r_C$ is
the response suppression function of category $C$, and $h_l$ is the number of
recommendations for customer $l$.

Now given new customers, we classify them into the above categories according to
their profiles and preference history.
As the response suppression function for each new customer,
we use the function of his/her corresponding category.
The original multicampaign model \cite{KM06} is very simple,
in the sense that the number of categories is just one.

\subsection{Formal Definition\label{sec:def}}

When we use the realistic response suppression function of Section~\ref{sec:R},
the fitness $F(M)$ is redefined in the following formula.
$$F(M) = \sum_{j=1}^k \sum_{i=1}^n w_j r_i(h_i) p_{ij} m_{ij},$$
where $h_i = \sum_{j=1}^k m_{ij}$.

Now we formally define a decision version of the multicampaign assignment problem
with the realistic response suppression function $\boldsymbol{r}$.
For convenience, we assume that $\boldsymbol{1}_n$ is the $1 \times n$ matrix $(1~1~\cdots~1)$.

\begin{Def}
{\em MCAP} 
$= \{ \langle \boldsymbol{w}, P, \boldsymbol{r}, \boldsymbol{b^*}, \boldsymbol{b_*}, t \rangle ~|~
\exists~M = (m_{ij}) \textrm{ such that } \boldsymbol{b_*} \le \boldsymbol{1}_n
M \le \boldsymbol{b^*} \textrm{ and }$ $F(M) \ge t \}$
\end{Def}

\subsection{Heuristic Algorithms\label{sec:heu}}

There have been a number of heuristic algorithms for solving MCAP \cite{DehC08,KM06,KM06-2}.
As stand-alone heuristics, the constructive assignment algorithm with $O(nk^2\log n)$-time complexity and
the iterative improvement heuristic with $O(nk \log n)$-time complexity were proposed in \cite{KM06}.
Recently, a novel heuristic using Lagrange multipliers with $O(nk^2)$-time complexity was devised \cite{KM06-2}.  
When combined with real-coded genetic algorithms \cite{Tsut01}, this algorithm could perform near-optimally.
There was also a recent heuristic algorithm of \cite{DehC08} based on particle swarm optimization \cite{KenE95}.
In summary, many kinds of heuristics for MCAP have been proposed and they showed good performance.
Most importantly, the practical version of the problem can also be solved 
by all these heuristic algorithms with the same complexity.
However, up to now nobody has shown whether or not MCAP is NP-complete.
In the next section, we focus on the computational complexity of MCAP.

\section{Computational Complexity of Multicampaign Assignment\label{sec:npc}}
Suppose there are several assignment or allocation problems,
some problems are in P (e.g., the optimal assignment problem \cite{Kuhn55}).
But, many of them are NP-complete \cite{GareCom79}
(e.g., the quadratic assignment problem \cite{Cela98},
the generalized assignment problem \cite{ChuB97}, and
the multiprocessor document allocation problem \cite{FriS97}).
In this section, we prove that the practical version of MCAP is NP-complete.
In the following extreme cases, MCAP is tractable, i.e., belongs to P.
\begin{itemize}
\item All the response suppression function are constant functions. Then the
problem can be solved optimally in $O(nk\log n)$ time.
If we only sort actual preferences of all the customers for each campaign $j$,
i.e., the set $\{p_{1j} r_1, p_{2j} r_2, \ldots, p_{nj} r_n \}$ for each $j$,
the optimal solution can be easily obtained.
\item There is no bound constraint, i.e., $\boldsymbol{b_*} = \boldsymbol{0}_k$
and $\boldsymbol{b^*} = n\boldsymbol{1}_k$. Then there exists an $O(nk \log k)$-time
algorithm to solve the problem to optimality.
In solving the problem, the most time-consuming task is to sort preferences of
each customer $i$ for all the campaigns, i.e., the set $\{p_{i1}, p_{i2}, \ldots, p_{ik} \}$ for each $i$.
\end{itemize}
However, in the general case, the problem becomes intractable by the following
theorem.

\begin{Thm} \label{thm:npc1}
{\em MCAP} is NP-complete.
Moreover, even when the campaign weight vector is ignored, i.e.,
$w_i=1$, the problem is still NP-complete.
\end{Thm}

{
\noindent {\sc Proof}:
To show that MCAP is in NP,
we let the campaign assignment matrix $M = (m_{ij})$ be the certificate.
For any instance of the problem, we can check whether $\boldsymbol{b_*} \le \boldsymbol{1}_n
M \le \boldsymbol{b^*}$ and $F(M) \ge t$ in polynomial time.

We show that 3-SAT $\le_P$ MCAP. Given a 3-CNF (conjunctive normal form)
formula $\phi$ over variables $x_1, x_2, \ldots, x_l$ with clauses
$C_1, C_2, \ldots, C_m$, each containing exactly three distinct literals,
the reduction algorithm constructs an instance
$\langle \boldsymbol{w}, P, \boldsymbol{r}, \boldsymbol{b^*}, \boldsymbol{b_*}, t \rangle$
of MCAP such that the formula $\phi$ is satisfiable if and only if there is a
campaign assignment matrix $M = (m_{ij})$ satisfying $\boldsymbol{b_*} \le
\boldsymbol{1}_n M \le \boldsymbol{b^*}$ and $F(M) \ge t$.
Without loss of generality, we make two assumptions about the formula $\phi$.\\
i) No clause contains both a variable and its negation.
Such a clause is always satisfied by any assignment of values to the variables.\\
ii) Each variable appears in at least one clause.
Otherwise it does not matter what value is assigned to the variable.

The reduction creates two customers, $u_i$ and $u_i'$, for each variable $x_i$ and
three customers, $s_j$, $s_j'$, and $s_j''$, for each clause $C_j$, i.e., $n = 2l+3m$.
We also create $k = l+m$ campaigns and each campaign corresponds to either one variable or one clause.
We label each campaign by either a clause or a variable; that is, the campaigns
are labelled $C_1, C_2, \ldots, C_m, x_1, x_2, \ldots, x_l$.
We construct $\boldsymbol{w}, P, \boldsymbol{r}, \boldsymbol{b^*}, \boldsymbol{b_*}$, and $t$ as follows:\\
-- The campaign weight vector $\boldsymbol{w}$ is set to $\boldsymbol{1}_n$.\\
-- The preference matrix $P = (p_{ij})$ is constructed as follows.
For each $i = 1, 2, \ldots, l$,
set $p_{u_ix_i} = p_{u_i'x_i} := 10^{m+i-1}$.
If literal $x_i$ appears in clause $C_j$, then set $p_{u_iC_j}$ to $10^{j-1}$.
If literal $\neg x_i$ appears in clause $C_j$, then set $p_{u_i'C_j}$ to $10^{j-1}$.
For each other variable or clause $y$, set $p_{u_iy} = p_{u_i'y} := 0$.
For each $j = 1, 2, \ldots, m$,
set $p_{s_jC_j} = p_{s_j'C_j} = p_{s_j''C_j} := 10^{j-1}$.
For each other variable or clause $y$, set $p_{s_jy} = p_{s_j'y} = p_{s_j''y} := 0$.\\
-- The response suppression function $\boldsymbol{r}$ is constructed as follows.
For each $s_j$, if $y = 1$, set $r_{s_j}(y)$ to 1. Otherwise set $r_{s_j}(y)$ to 0.
Set $r_{s_j'} = r_{s_j''} := r_{s_j}$.
For each $u_i$, let $\alpha_i$ be $1 + \sum_{C_j;P_{u_iC_j}>0} 1$.
If $y = \alpha_i$, set $r_{u_i}(y)$ to 1. Otherwise set $r_{u_i}(y)$ to 0.
For each $u_i'$, let $\alpha_i'$ be $1 + \sum_{C_j;P_{u_i'C_j}>0} 1$.
If $y = \alpha_i'$, set $r_{u_i'}(y)$ to 1. Otherwise set $r_{u_i'}(y)$ to 0.\\
-- The upper bound capacity constraint vector $\boldsymbol{b^*}$ is as follows.
Set $b^{C_1} = b^{C_2} = \cdots = b^{C_m} := 4$ and $b^{x_1} = b^{x_2} = \cdots = b^{x_l} := 1$, i.e.,
$\boldsymbol{b^*} = (4~4~\cdots~4~1~1~\cdots~1)$.\\
-- The lower bound capacity constraint vector $\boldsymbol{b_*}$
is equal to $\boldsymbol{b^*}$.\\
-- The value $t$ is set to the number in base 10 
represented by the digit vector $\boldsymbol{b^*}$, where the clause campaigns
$C_1, C_2, \ldots, C_m$ stand for the $m$ least significant digits, and the
variable campaigns $x_1, x_2, \ldots, x_l$ stand for the $l$ most significant digits,
i.e., $t = \sum_{j=1}^k b^j 10^{j-1} = 4 \sum_{i=1}^m 10^{i-1} + \sum_{j=1}^l 10^{m+j-1}$.

The reduction can be performed in polynomial time. 
Figure~\ref{fig:ex} shows the reduction of an example formula with $3$ variables and $4$
clauses. This will help to understand the reduction algorithm.

\begin{figure}[tb]
\hspace*{1.5cm} \psfig{figure=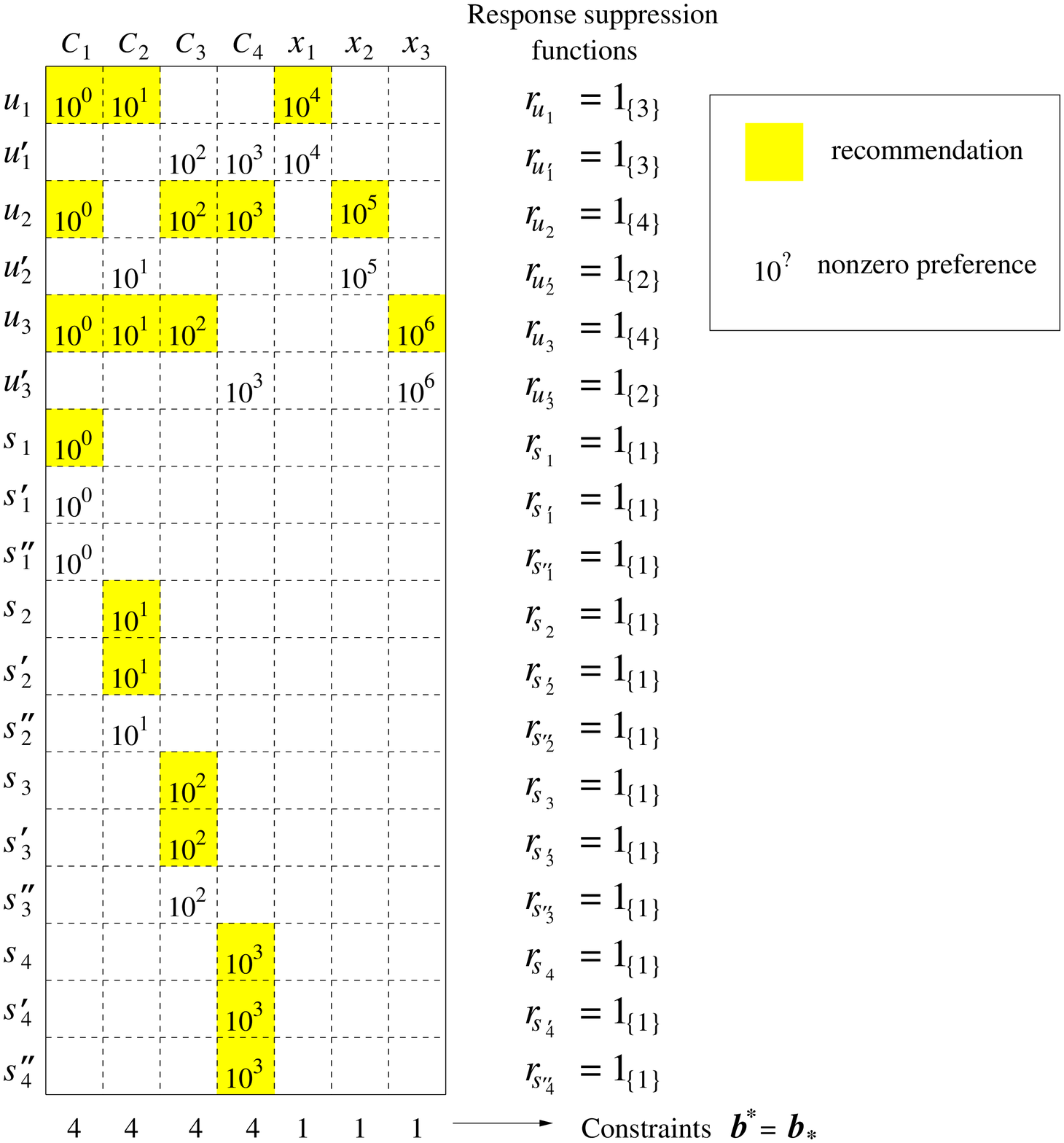,width=4.45in,angle=0}\\
\hspace*{1.5cm} $\phi = (x_1 \vee x_2 \vee x_3) \wedge (x_1 \vee \neg x_2 \vee
x_3) \wedge
(\neg x_1 \vee x_2 \vee x_3) \wedge (\neg x_1 \vee x_2 \vee \neg x_3)$\\
\hspace*{1.5cm} {\small * $1_{A}$ is the indicator function, i.e., $1_{A}(x) =
1$ if $x \in
A$, otherwise, $1_{A}(x)=0$.}
\caption{\label{fig:ex}An example of the reduction 3-SAT $\le_p$ MCAP}
\end{figure}

Now we show that
the 3-CNF formula $\phi$ is satisfiable if and only if there is a
campaign assignment matrix $M = (m_{ij})$ satisfying 
$\boldsymbol{1}_n M = \boldsymbol{b^*} (= \boldsymbol{b_*})$ and $F(M) \ge t$.
First, suppose that $\phi$ has a satisfying assignment.
For $i = 1,2,\ldots,l$, if $x_i = 1$ in this assignment, then set $m_{u_ix_i}$
to 1. Otherwise, set $m_{u_i'x_i}$ to 1.
Having either $m_{u_ix_i}=1$ or $m_{u_i'x_i}=1$, but not both, for all $i$,
and setting $m_{s_jx_i} = m_{s_j'x_i} = m_{s_j''x_i}$ to 0 for all $i$ and $j$,
we see that for each variable-labeled campaign, the sum of recommendations must be 1,
which matches the values of $\boldsymbol{b^*}$.
Because each clause is satisfied, there is some literal in the clause with the
value one. Then, each campaign labeled by a clause $C_j$ has at least one $u_i$
or $u_i'$ customer that has positive preference for campaign $C_j$ and
is recommended by campaign $x_i$. In fact, 1, 2, or 3 literals may be
$1$ in each clause, and so, among the $u_i$ and $u_i'$ customers,
each clause-labeled campaign $C_j$ has 1, 2, or 3 customers
that have positive preference for campaign $C_j$ and
are recommended by campaign $x_i$.
For each $u$ among such customers, set $m_{uC_j}$ to 1.
We achieve four recommendations in each campaign labeled by clause $C_j$ by
setting $m_{sC_j}$ to 1 only for each customer $s$ in the appropriate nonempty 
subset of customers $\{s_j, s_j', s_j''\}$.
We have matched $\boldsymbol{b^*}$ ($=\boldsymbol{b_*}$) in all campaigns.
If $M$ is assigned as described above, then $r_i(h_i) = 1$ for each customer $i$ and
$p_{ij}=10^{j-1}$ whenever $m_{ij}=1$.
Hence, $F(M) = \sum_{i=1}^{n} \sum_{j=1}^k w_j r_i(h_i) p_{ij} m_{ij}
= \sum_{i=1}^{n} \sum_{j=1}^k p_{ij} m_{ij}
= \sum_{j=1}^{k} \sum_{i=1}^n 10^{j-1} m_{ij}
= \sum_{j=1}^{k} 10^{j-1} b^j = t$.

Now, suppose that there is a campaign assignment matrix $M = (m_{ij})$ such that
$\boldsymbol{1}_n M = \boldsymbol{b^*} (= \boldsymbol{b_*})$ and $F(M) \ge t$.
Since $r_i(h_i) \le 1$ for each customer $i$ and $p_{ij} \le 10^{j-1}$,
$F(M) = \sum_{i=1}^{n} \sum_{j=1}^k w_j r_i(h_i) p_{ij} m_{ij} \le
\sum_{i=1}^{n} \sum_{j=1}^k p_{ij} m_{ij} \le \sum_{j=1}^{k} 10^{j-1} b^j = t$.
Hence, $F(M)$ is equal to $t$.
The following important properties about the matrix $M$ hold.\\ 
{\bf Property 1} For each customer $i$ and each campaign $j$, if $m_{ij} = 1$, 
then $p_{ij} = 10^{j-1}$.\\
($\because$ Suppose that $p_{ij} \ne 10^{j-1}$, i.e., $p_{ij} = 0$.
Then, $F(M) = \sum_{i=1}^{n} \sum_{j=1}^k w_j r_i(h_i) p_{ij} m_{ij} \le
\sum_{i=1}^{n} \sum_{j=1}^k p_{ij} m_{ij} < \sum_{j=1}^{k} 10^{j-1} \sum_{i=1}^{n} m_{ij} = t$.
Hence, $F(M) < t$. This is a contradiction.)\\
{\bf Property 2} If  $m_{ij} = 1$, then $r_i(h_i)=1$.\\
($\because$ Suppose that $r_i(h_i) \ne 1$, i.e., $r_i(h_i) =0$.
Then, $F(M) = \sum_{i=1}^{n} \sum_{j=1}^k w_j r_i(h_i) p_{ij} m_{ij} <
\sum_{i=1}^{n} \sum_{j=1}^k p_{ij} m_{ij} \le t$.
Hence, $F(M) < t$. This is a contradiction.)\\
{\bf Property 3} If $m_{ij} = 1$, then $m_{iq} = 1$ for every campaign $q$ 
satisfying $p_{iq} > 0$.\\
($\because$ Since $r_i(h_i)=1$ by Property 2, $h_i(= \sum_{j=1}^{k} m_{ij})
= \alpha_i$, i.e., $h_i$ is the same as
the number of $j$'s such that $p_{ij}>0$. Also, if $p_{ij}=0$, $m_{ij}=0$
by Property 1. So, if there exists campaign $q$ such that 
$p_{iq} > 0$ and $m_{iq} = 0$, $h_i$ cannot be $\alpha_i$.)\\
Consequently, $M$ chooses all elements that have positive preference 
in one row, or choose nothing.

For each $i = 1, 2, \ldots, l$, either $m_{u_ix_i} = 1$ or $m_{u_i'x_i} = 1$,
but not both, for otherwise the campaigns labeled by variables would not sum to 1.
If $m_{u_ix_i} = 1$, set $x_i$ to 1. Otherwise set $x_i$ to 0.
We claim that every clause $C_j$, for $j = 1, 2, \ldots, m$, is satisfied by
this statement. To prove this claim, note that to achieve four recommendations in
the campaign labeled by $C_j$, for at least one $u_i$ or $u_i'$, $m_{u_iC_j} = 1$ or 
$m_{u_i'C_j} = 1$, since the contributions of customers $s_j$, $s_j'$, and
$s_j''$ together sum to at most 3. If $m_{u_iC_j} = 1$, then $m_{u_ix_i} = 1$
by Property 3 and $p_{u_iC_j} > 0$ by Property 1, which means that the literal $x_i$ appears in $C_j$. 
Since we have set $x_i = 1$ when $m_{u_ix_i} = 1$, clause $C_j$ is satisfied. 
If $m_{u_i'C_j} = 1$, then $m_{u_i'x_i} = 1$ and $p_{u_i'C_j} > 0$,
and so the literal $\neg x_i$ appears in $C_j$. Since we have set $x_i = 0$
when $m_{u_i'x_i} = 1$, clause $C_j$ is satisfied.
Thus, all clauses of formula $\phi$ are satisfied.\\
Q.E.D.
}

\section{Concluding Remarks}\label{sec:con}
In this paper, we suggested a practical version of the multicampaign assignment problem (MCAP).
Under the practical multicampaign model, we also presented how to determine a realistic response suppression function
instead of existing artificial functions of \cite{DehC08,KM06}.

In the case that the number of campaigns is fixed, 
the optimal campaign assignment matrix can be found in polynomial time
using the dynamic programming method. However, in general, the method is intractable.
The MCAP could also be solved near-optimally in a reasonable time by novel heuristics
proposed in our prior work \cite{KM06-2}. However, its intractability has been left open.
In this paper, in confirmation of the intractability,
we provided an elaborate proof of that the problem is NP-complete.
The NP-completeness of the MCAP justifies the use of various heuristics for the problem as in \cite{DehC08,KM06,KM06-2}.


\subsection*{Acknowledgments}
The authors would like to thank Prof. Vic Rayward-Smith
for his valuable suggestions in improving this paper.

\bibliographystyle{plain}
\bibliography{ref}

\end{document}